\documentclass{aptpub}
\usepackage{amsmath}
\usepackage{graphicx}

\authornames{D. A. Kessler, S. Medalion, E. Barkai}
\shorttitle{The Distribution of the Area under a Bessel Excursion and its Moments}

\begin{document}
\title{The Distribution of the Area under a Bessel Excursion and its Moments}

\authorone[Bar-Ilan University]{David A. Kessler}
\addressone{Dept. of Physics, Bar-Ilan University, Ramat-Gan IL52900 Israel\\Email address: kessler@dave.ph.biu.ac.il}
\authortwo[Bar-Ilan University]{Shlomi Medalion}
\addresstwo{Dept. of Physics and Inst. of Nanotechnology and Advanced Materials, Bar-Ilan University, Ramat-Gan IL52900 Israel\\Email address: shlomed.uni@gmail.com}
\authorthree[Bar-Ilan University]{Eli Barkai}
\addressthree{Dept. of Physics and Inst. of Nanotechnology and Advanced Materials, Bar-Ilan University, Ramat-Gan IL52900 Israel\\Email address: Eli.Barkai@biu.ac.il}
\begin{abstract}
A Bessel excursion is a Bessel process that begins at the origin and first returns there at some given time $T$.  We study the distribution of the area under such an excursion, which recently found application in the context of laser cooling.  The area $A$ scales with the time as $A \sim T^{3/2}$, independent of the dimension, $d$,  but the functional form of the distribution does depend on $d$.  We demonstrate that for $d=1$, the distribution reduces as expected to the distribution for the area under a Brownian excursion, known as the Airy distribution,  deriving a new expression for the Airy distribution in the process. We show that the distribution is symmetric in $d-2$, with nonanalytic behavior at $d=2$.  We calculate the first and second moments of the distribution, as well as a particular fractional moment.  We also analyze the analytic continuation from $d<2$ to $d>2$. In the limit where $d\to 4$ from below, this analytically continued distribution is described by a one-sided L\'evy $\alpha$-stable distribution with index $2/3$ and a scale factor proportional to $[(4-d)T]^{3/2}$.
\end{abstract}
\section{Introduction}
The Brownian random walk has been the subject of countless works, and many different aspects of the walk have been studied and found to have important applications.  Among these is the area under a Brownian excursion, a  Brownian random walk that starts at the origin and first returns to the origin at time $T$. For a review of Brownian excursions and related problems, see Ref. \cite{review}. The distribution of the area has been calculated analytically and been entitled the Airy distribution, as it involves in various ways the Airy function~\cite{Darling,Louchard}.  A natural generalization of the Brownian random walk is the Bessel process~\cite{Ito}, corresponding to the Langevin equation
\begin{equation}
\dot{x} = -\frac{DU_0}{x} + \eta
\end{equation}
where $\eta$ is a Gaussian white noise satisfying
\begin{equation}
\langle \eta \rangle = 0; \qquad  \langle \eta(t)\eta(t') \rangle = 2D\delta(t-t')
\end{equation}
The drift term $-DU_0/x$ can be thought as the limiting case of a regularized drift term $-DU_0 x/(x_0^2 + x^2)$ for vanishingly small $x_0$.
 In addition, the Bessel process for  integer $U_0\le 0$ corresponds to the fluctuations in the distance to the origin of a Brownian process in $d=1-U_0$ 
 dimensions. In parallel with the Brownian excursion, one can define what one has been called~\cite{Hu} the Bessel excursion, where the walker first returns 
 to the origin at  time $T$.  This leads one to consider the area under the Bessel excursion, whose distribution we shall call the Bessel distribution.  This 
 distribution has been shown~\cite{PRL3} to play a vital role in determining the position of atoms in an optical lattice undergoing Sisyphus cooling (with
  $U_0 >0)$.  In this paper, we shall derive a formula for the Bessel distribution and for its Laplace transform.  We shall also consider the moments of the 
 distribution, in particular giving a simple explicit formula for the first moment.  We show that the distribution is symmetric in $d-2$, with nonanalytic behavior 
at $d=2$.  We also study the analytic continuation of the distribution from $d<2$ to the region $d>2$.  This analytically continued distribution  exhibits critical 
behavior in the vicinity of $U_0=-3$, corresponding to $d=4$.  Here the distribution is described to leading order by a fat-tailed L\'evy $\alpha$-stable 
distribution on an inner scale of $[(4-d)T]^{3/2}$, cut off at areas of order $T^{3/2}$.

The order of the paper is as follows.  In Section \ref{secDist}, we derive an expression for the Bessel Distribution, first in Laplace space and subsequently in real space. In Section \ref{secAiry}, we discuss the connection to the Airy distribution, corresponding to the limit $U_0=0$.  In the following section, we discuss the large $k$ asymptotics of the quantities $\lambda_k$ and $d_k$ which enter into our expression for the Bessel Distribution.  We then turn to a calculation of the zeroth, first and second moments of the Bessel Distribution. In the following section, we show how the calculation simplifies in the Airy Distribution limit, allowing for a straightforward calculation of arbitrary integer moments.  In Section \ref{secU0m3}, we deal with the analytical continuation of the distribution above $d=2$ (below $U_0=-1$), examining in detail the limit $U_0+3\ll 1$, corresponding to $4-d \ll 1$. In the penultimate section, we produce a closed form expression for the $(U_0+1)/3$rd moment.  We then conclude with a summary and some observations.

\section{The Bessel Distribution\label{secDist}}
Our first goal is to calculate the Bessel Distribution, $P(A,T)$, where $A\equiv \int_0^T x(t)dt$ is the area under the excursion $x(t)$. To do this we shall employ a generalization~\cite{Carmi} to Fokker-Planck equations of the Feynman-Kac formula~\cite{Kac} for the Laplace transform of the distribution with respect to the area $A$, $\tilde{P}(s,T)\equiv \int_0^\infty P(A,T) e^{-sA} dA$.  The fundamental object we need in order to calculate $\tilde{P}(s,T)$ is the area-weighted propagator 
\begin{equation}
G_s(x,x_0;T) = \sum_{p} e^{-s A_p} G(x,x_0;T)
\end{equation}
where the sum is over all positive paths $p$ (paths which do not cross $x=0$) $p$ starting at $x_0$ at $t=0$ and arriving at $x$ at time $T$, $A_p$ is the area under the corresponding path and $G$ is the standard propagator, satisfying the Fokker-Planck equation
\begin{equation}
D\left[\frac{\partial^2}{\partial x^2} G + \frac{\partial}{\partial x} \left(\frac{U_0}{x}G \right)\right]  = \frac{\partial}{\partial T}G; \qquad G(x,x_0;0)=\delta(x-x_0).
\label{FP}
\end{equation}
The generalized Feynman-Kac equation for $G_s$ is:
\begin{equation}
D\left[\frac{\partial^2}{\partial x^2} G_s + \frac{\partial}{\partial x} \left(\frac{U_0}{x}G_s \right)\right] - sx G_s = \frac{\partial}{\partial T}G_s
\label{FK}
\end{equation}
with the initial condition $G_s(x,x_0;0) = \delta(x-x_0)$.  For $U_0=0$, we get the original Feynman-Kac equation corresponding to Brownian functionals. Since we are analyzing excursions, which  do not cross the origin in the interval $(0,T)$, we have to
impose the condition $G_s(0,x_0;T) = 0$.  In terms of this weighted propagator, the Laplace-transformed distribution for the area under excursions, which by definition start and end on the origin, is given by
\begin{equation}
\widetilde{P}(s,T) = \lim_{x=x_0\to 0} \frac{G_s(x,x_0;T)}{G_0(x,x_0;T)}
\label{Ptilde}
\end{equation}
where the denominator ensures the correct normalization, which is well-defined for any finite $x=x_0$.  

This calculation for the case $U_0=0$, i.e., Brownian excursions, was performed by Majumdar and Comtet~\cite{Majumdar}. The present calculation proceeds along the same lines but the singular and non-Hermitian nature of the $U_0$ term requires special care.   Our first step is to perform a similarity transformation to transform the Fokker-Planck operator in Eq. (\ref{FK}) to a Schr\"odinger operator.  We define 
\begin{equation}
G_s(x,x_0;T) = \left(\frac{x}{x_0}\right)^{-U_0/2} K_s(x,x_0;T)
\label{Gs}
\end{equation}
$K_s$ then satisfies the equation
\begin{equation}
D\left(\frac{\partial^2}{\partial x^2} K_s - U_0 \frac{U_0+2}{4x^2} K_s \right) - sx K_s = \frac{\partial}{\partial T}K_s 
\label{Schr_s}
\end{equation}
with initial condition $K_s(x,x_0;0)=\delta(x-x_0)$. It will be important for us later that $K_s$ is the imaginary-time propagator of the Schr\"odinger operator 
\begin{equation}
\hat{H} = -D \frac{\partial^2}{\partial x^2} + \underbrace{D U_0 \frac{U_0+2}{4x^2}  + sx }_{V_\textit{eff}(x)}
\label{Veff}
\end{equation}
corresponding to an ``effective potential", $V_\textit{eff}(x)$, consisting of a linear potential with a centrifugal barrier.  As such, $K_s$ satisfies
\begin{equation}
\hat{H}K_s + \frac{\partial K_s}{\partial t} = \delta(x-x_0) \delta(t)
\end{equation}

We can construct $K_s$ via an eigenvalue expansion, scaling out $s$ and $D$ in the process.  Assuming the spectrum is discrete, as we will verify momentarily, we define the normalized eigenvectors, $\phi_k$ and eigenvalues, $\lambda_k$ of the rescaled Hamiltonian,  $\bar{H}$, through
\begin{equation}
\bar{H}\phi_k \equiv -\frac{\partial^2}{\partial x^2} \phi_k + \left(U_0 \frac{U_0+2}{4x^2} + x \right)\phi_k = \lambda_k \phi_k
\label{Schro}
\end{equation}
with 
\begin{equation}
\int_0^\infty \phi_k^2(x) = 1.
\end{equation}
We need to consider carefully the boundary conditions.
The effective potential of the Schr\"odinger problem, $V_\textit{eff}(x)$ grows as $x\to \infty$, given that $s>0$, and so $\phi_k(x) \to 0$ as $x\to \infty$. As we wish to exclude paths that return to the origin before time $T$, we enforce absorbing boundary conditions at some small $x=\epsilon$, setting $\phi_k(\epsilon)= 0$ and at the end take $\epsilon \to 0^+$. This procedure is crucial due to the singular nature of the origin for the Bessel process. (For a general discussion of boundary conditions for Bessel processes, see Ref. \cite{PREBessel}.) For small $x$, the general solution of Eq. (\ref{Schro}) is given by
\begin{equation}
\phi_k \approx E_k(\epsilon) x^{1/2 + |\alpha|} + F_k(\epsilon) x^{1/2 - |\alpha|}
\label{eqbc0}
\end{equation}
where we have explicitly noted the $\epsilon$ dependence of the coefficients and introduced 
\begin{equation}
\alpha \equiv \frac{U_0 + 1}{2} .
\end{equation}
In order to ensure that $\phi_k(\epsilon) = 0$, we have to have that $F_k/E_k \sim {\cal{O}}(\epsilon^{2|\alpha|})$.  Thus, in the limit $\epsilon \to 0^+$,
the $F_k$ term vanishes and the effective boundary condition becomes that 
\begin{equation}
\phi_k \sim x^{1/2 + |\alpha|}; \qquad x\ll 1.
\label{eqbc}
\end{equation}
 With these boundary conditions, it is clear that the spectrum is indeed discrete.
Given the $\phi_k$, we have
\begin{equation}
K_s(x,x_0;T) = \sum_k \beta \phi_k(\beta x) \phi_k(\beta x_0)  e^{-\gamma \lambda_k T} ,
\label{eqKs}
\end{equation}
with
\begin{equation}
\beta = \left(\frac{s}{D}\right)^{1/3} ;\qquad \gamma= D^{1/3} s^{2/3} .
\end{equation}
Thus, using Eq. (\ref{Gs}),
\begin{equation}
G_s(x,x_0;T) = \sum_k \beta \phi_k(\beta x) \phi_k(\beta x_0) \left(\frac{x_0}{x}\right)^{U_0/2} e^{-\gamma \lambda_k T} .
\end{equation}
According to Eq. (\ref{Ptilde}), we also need the $s=0$ propagator $G_0$.  For this, we return to the unscaled equation, Eq. (\ref{Schr_s}), which is the equation for the propagator for a free particle with a centrifugal barrier. Here there is no linear component of the potential, so to keep the spectrum discrete we adopt the standard artifice of placing an infinite potential barrier at $x=L$, so that the eigenfunctions $\phi_k^0$ vanish at $L$, and take $L\to\infty$ at the end. Then,
\begin{equation}
K_0(x,x_0;T)=\sum_k \phi^0_k(x)\phi^0_k(x_0)  e^{-Dk^2 T} ,
\label{eqK0}
\end{equation}
where due to  the boundary condition, Eq. (\ref{eqbc}),
\begin{equation}
\phi^0_k(x) = \left(\frac{\pi}{L}\right)^{1/2}\sqrt{kx} J_{|\alpha|}(k x) ,
\end{equation}
and $J_{|\alpha|}$ is a Bessel function.
Then, for $L \gg 1$, the sum in Eq. (\ref{eqK0}) can be replaced by an integral, and we have
\begin{align}
K_0(x,x_0;T) &=  \int_0^\infty \frac{dk L}{\pi} \frac{\pi}{L} k\sqrt{x x_0} J_{|\alpha|}(k x) J_{|\alpha|}(k x_0) e^{-Dk^2 T} \nonumber\\
&= \frac{\sqrt{x x_0}}{2DT} e^{-(x^2+x_0^2)/(4DT)} I_{|\alpha|}\left(\frac{x x_0}{2DT}\right) ,
\label{K0}
\end{align}
where $I_{|\alpha|}$ is a modified Bessel function of the first kind,
and so, again using Eq. (\ref{Gs}),
\begin{equation}
G_0(x,x_0;T) = \left(\frac{x_0}{x}\right)^{U_0/2}\frac{\sqrt{x x_0}}{2DT} e^{-(x^2+x_0^2)/(4DT)} I_{|\alpha|}\left(\frac{x x_0}{2DT}\right) .
\label{G0}
\end{equation}

To take the requisite $x=x_0\to 0$ limit, we need to explicitly parametrize the small $x$ behavior of $\phi_k$, writing, based on Eqs. (\ref{eqbc0},\ref{eqbc}),
\begin{equation}
\phi_k(x) \approx d_k x^{1/2 + |\alpha|}
\label{dk}
\end{equation}
where $d_k\equiv \lim_{\epsilon \to 0^+} E_k(\epsilon)$.
In addition, $I_{|\alpha|}(x) \approx (x/2)^{|\alpha|} /\Gamma(|\alpha|+1)$, and so, using Eqs. (\ref{Ptilde}), (\ref{G0}) and (\ref{dk}),
\begin{equation}
\widetilde{P}(s,T) = 2^{2|\alpha| + 1} \Gamma(|\alpha|+1)T^{|\alpha|+1}D^{|\nu|/2+1/3} s^{|\nu|+2/3} \sum_k d_k^2 e^{-D^{1/3}s^{2/3} \lambda_k T}
\label{Ptilde1}
\end{equation}
where for convenience we have introduced
\begin{equation}
\nu\equiv \frac{2\alpha}{3}= \frac{U_0+1}{3} .
\end{equation}
The first remarkable conclusion is that $\widetilde{P}(s,T)$ (and so also $P(A,T)$) is an even function of $\alpha$.  This is ultimately a result of the fact that $U_0$ in Eq. (\ref{Schr_s}) only appears in the combination $U_0(U_0+2) = 4\alpha^2 - 1$, which is even in $\alpha$. In particular, it means that the case
$d=3$ corresponding to $U_0 = -2$, $\alpha=-1/2$ gives the same distribution as $d=1$, namely $U_0=0$, $\alpha=1/2$.  In addition, we see that
 $\widetilde{P}$ is a function only of the dimensionless scaling variable 
\begin{equation}
\hat{s} \equiv D^{1/2}T^{3/2} s \equiv A_0 s, 
\end{equation}
which since $s$ scales as the inverse of the area, implies that the area scales as $T^{3/2}$, for all $U_0$.  The fact that the scaling behavior is independent of $U_0$, while the functional form of the distribution does change with $U_0$, is a unique consequence of the marginal nature of the $1/x$ bias of the Bessel excursion.
In the limit $U_0\to 0$, $\alpha=1/2$, $\nu=1/3$, and $d_k^2=1$ (as can be seen from the direct solution of the Schr\"odinger equation in terms of 
Airy functions), so that our formula reduces to
\begin{equation}
\widetilde{P}(s,T) = 2\sqrt{\pi}  \hat{s} \sum_k  e^{-\lambda_k \hat{s}^{2/3}}
\label{Ptilde0}
\end{equation}
which corresponds to the known result~\cite{Darling,Louchard}, where traditionally $D=1/2$ as is appropriate for a random walk. 

\begin{figure}
\includegraphics[width=0.7\textwidth]{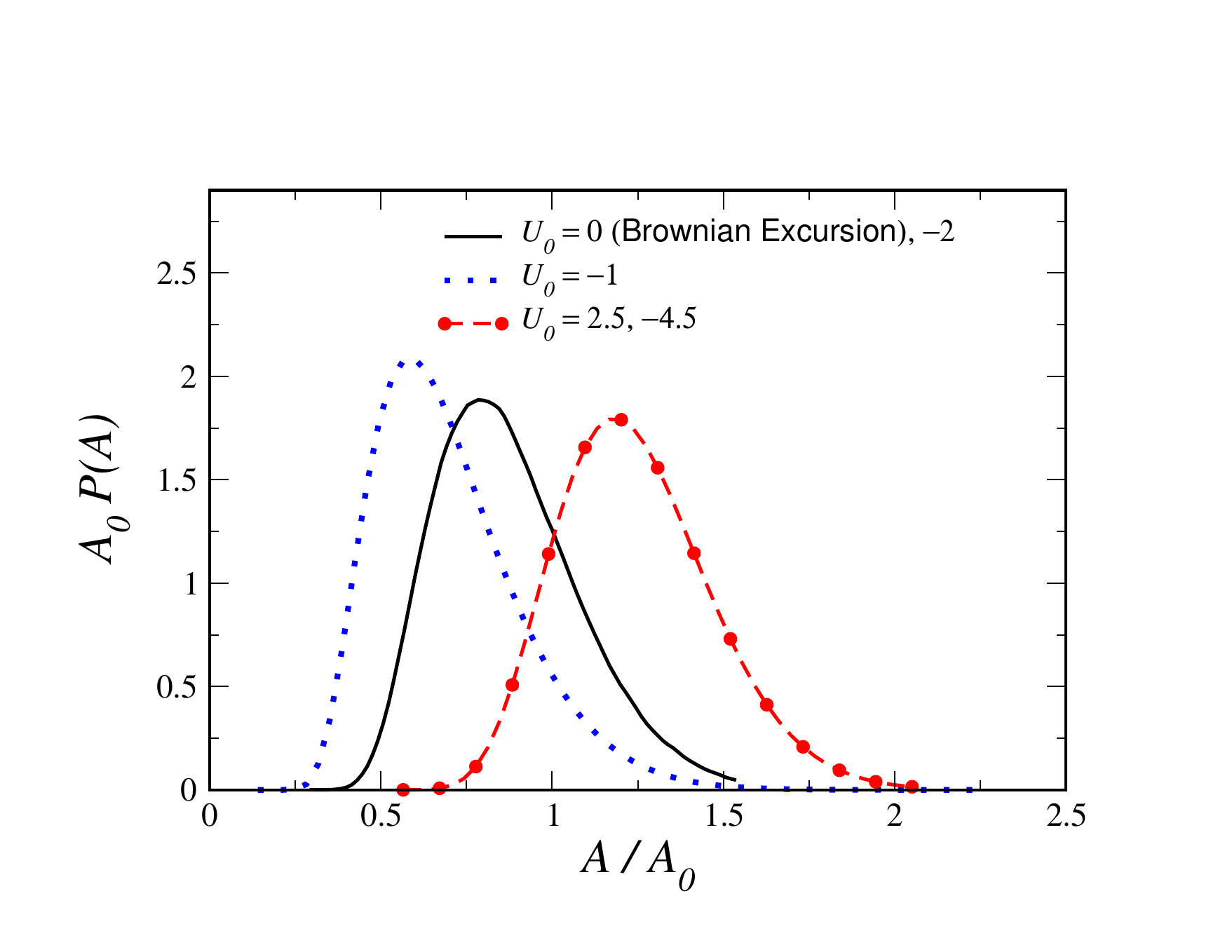}
\caption{The Bessel Distribution of the area $A$ under the Bessel excursion for the cases $U_0=-1$, $0$, (equivalent to $U_0=-2$), and $2.5$ (equivalent to $U_0=-4.5$). The case $U_0=0$ is the classic Airy Distribution for the area under a Brownian excursion.  The case $U_0=-1$ corresponds to the radial component of $d=2$ Brownian motion.  When comparing to standard treatments, it should be remembered that it is customarily assumed that $D=1/2$, $T=1$ so that $A_0=1/\sqrt{2}$.}
\label{figP}
\end{figure}

Likewise, $A_0 P(A,T)$ is a function only of the scaling variable $\hat{A} = A/A_0$. It is possible to calculate this by performing the inverse Laplace transform of $\widetilde{P}$,  expanding term by term in powers of $\hat{s}^{2/3}$.  Using the fact that the inverse Laplace transform of $s^p$ is $x^{-(1+p)}/\Gamma(-p)$, we get, using Maple to resum the series,
\begin{align}
{\cal L}^{-1}\Big[s^{\nu + 2/3} &e^{-as^{2/3}}\Big] =\nonumber\\{} &-\frac{1}{\pi x^{\nu+5/3}}\Bigg[\Gamma\left(\frac{5}{3}+\nu\right)\sin\left(\pi\frac{2+3\nu}{3}\right){}_2F_2\left(\frac{4}{3}+\frac{\nu}{2},\frac{5}{6}+\frac{\nu}{2};\frac{1}{3},\frac{2}{3};-\frac{4a^3}{27 x^2}\right)\nonumber\\
& {}\qquad\qquad\quad -\frac{a}{x^{2/3}} \Gamma\left(\frac{7}{3}+\nu\right)\sin\left(\pi\frac{4+3\nu}{3}\right){}_2F_2\left(\frac{7}{6}+\frac{\nu}{2},\frac{5}{3}+\frac{\nu}{2};\frac{2}{3},\frac{4}{3};-\frac{4a^3}{27 x^2}\right) \nonumber\\
& {}\qquad\qquad\quad + \frac{1}{2}\left(\frac{a }{x^{2/3}}\right)^2 \Gamma\left(3+\nu\right)\sin\left(\pi\nu\right){}_2F_2\left(2+\frac{\nu}{2},\frac{3}{2}+\frac{\nu}{2};\frac{4}{3},\frac{5}{3};-\frac{4a^3}{27 x^2}\right)\Bigg]
\label{ilt}
\end{align}
where ${\cal L}^{-1}$ denotes the inverse Laplace transform.
Thus,
\begin{align}
P(A,T) &= -\frac{\Gamma(1+|\alpha|)}{2\pi A}\left(\frac{4}{\hat{A}^{2/3}}\right)^{|\alpha|+1}\times\nonumber\\
&\qquad \sum_k d_k^{2}\Bigg[\Gamma\left(\frac{5}{3}+|\nu|\right)\sin\left(\pi\frac{2+3|\nu|}{3}\right){}_2F_2\left(\frac{4}{3}+\frac{|\nu|}{2},\frac{5}{6}+\frac{|\nu|}{2};\frac{1}{3},\frac{2}{3};-\frac{4\lambda_k^3}{27 \hat{A}^2}\right)\nonumber\\
& {}\qquad\qquad -\frac{ \lambda_k}{\hat{A}^{2/3}} \Gamma\left(\frac{7}{3}+|\nu|\right)\sin\left(\pi\frac{4+3|\nu|}{3}\right){}_2F_2\left(\frac{7}{6}+\frac{|\nu|}{2},\frac{5}{3}+\frac{|\nu|}{2};\frac{2}{3},\frac{4}{3};-\frac{4\lambda_k^3}{27 \hat{A}^2}\right) \nonumber\\
& {}\qquad\qquad + \frac{1}{2}\left(\frac{\lambda_k }{\hat{A}^{2/3}}\right)^2 \Gamma\left(3+|\nu|\right)\sin\left(\pi|\nu|\right){}_2F_2\left(2+\frac{|\nu|}{2},\frac{3}{2}+\frac{|\nu|}{2};\frac{4}{3},\frac{5}{3};-\frac{4\lambda_k^3}{27 \hat{A}^2}\right)\Bigg]
\end{align}
which is what we call the Bessel Distribution.  This is graphed, using numerically computed values of the $\lambda_k$ and $d_k$, in Fig. \ref{figP} for the cases $U_0=-1$, $0$, and $2.5$, where by  symmetry, the distributions in the latter two cases are identical to those for $U_0=-2$, $-4.5$, respectively. We see that for $U_0>-1$, as $U_0$ increases the distribution shifts to the right, since the only paths that survive the increasing inward drift are those that wandered far from the origin, where the drift is weaker.  By symmetry, as $U_0$ decreases past $U_0=-1$, the distribution also shifts to to the right, despite the increasing bias away from the origin. 

\section{Relationship to the Airy Distribution\label{secAiry}}
As we showed above in Eq. (\ref{Ptilde0}), the Laplace transform $\widetilde{P}(s,T)$ goes over to the transform of the Airy distribution in the limit $U_0\to 0$.
To show the same for $P(A,T)$ itself is a little more involved.  We first consider the one-sided L\'evy $\alpha$-stable distribution with index $2/3$, whose Laplace transform is given by 
\begin{equation}
\tilde{L}_{2/3,1} (s) \equiv \int_0^\infty dx L_{2/3,1}(x) e^{-sx} = e^{-s^{2/3}}
\end{equation}
In standard treatments~\cite{Zolotarev}, this is given in real-space in terms of the Whittaker $W$ function:
\begin{equation}
L_{2/3,1}(x) = \frac{\sqrt{3/\pi}}{x}e^{-2/(27x^2)} W\left(\frac{1}{2},\frac{1}{6},\frac{4}{27x^2}\right)
\end{equation}
Our formula, Eq. (\ref{ilt}), with  $\nu=-2/3$, $a=1$ gives
\begin{align}
L_{2/3,1}(x) = \frac{\sin\frac{2\pi}{3}}{\pi x^{5/3}} &\Bigg[   \Gamma\left(\frac{5}{3}\right){}_2F_2\left(\frac{5}{6},\frac{4}{3};\frac{2}{3},\frac{4}{3};-\frac{4}{27 x^2}\right) \nonumber\\
&{} \quad + \left(\frac{1 }{2x^{2/3}}\right) \Gamma\left(\frac{7}{3}\right){}_2F_2\left(\frac{5}{3},\frac{7}{6};\frac{4}{3},\frac{5}{3};-\frac{4}{27 x^2}\right)\Bigg]
\end{align}
reproducing a formula presented in Ref. \cite{Penson}.
Both the ${}_2F_2$ factors are now degenerate, since one of the upper indices equals a lower index, and they reduce to ${}_1F_1$,  a confluent hypergeometric function, i.e. a Whittaker $M$ function~\cite{AbS}:
\begin{align}
{}_2F_2\left(\frac{5}{6},\frac{4}{3};\frac{2}{3},\frac{4}{3};-y\right)  = &{}_1F_1\left(\frac{5}{6};\frac{2}{3};-y\right) 
=e^{-y/2}y^{-1/3} M_{1/2,-1/6}(y) ,\nonumber\\
{}_2F_2\left(\frac{5}{3},\frac{7}{6};\frac{4}{3},\frac{5}{3};-y\right)  = &{}_1F_1\left(\frac{7}{6};\frac{4}{3};-y\right)
 =e^{-y/2}y^{-2/3} M_{1/2,1/6}(y) .
\end{align}
Putting this all together, with $y=4/(27x^2)$, we indeed get the expected Whittaker $W$ function~\cite{AbS}:
\begin{align}
L_{2/3,1}(x) &= \frac{\sqrt{3}}{2\pi x^{5/3}y^{1/3}}e^{-y/2}\left[\Gamma\left(\frac{5}{3}\right)M_{1/2,-1/6}(y) + \frac{3}{2^{4/3}}\Gamma\left(\frac{7}{3}\right)
M_{1/2,1/6}(y)\right] \nonumber\\
&= \frac{\sqrt{3/\pi}}{x} e^{-2/(27x^2)}W\left(\frac{1}{2},\frac{1}{6},\frac{4}{27x^2}\right) .
\end{align}
Alternatively, $L_{2/3,1}(x)$ can be written in terms of the Kummer $U$ function~\cite{AbS}:
\begin{equation}
L_{2/3,1}(x) = \frac{2^{4/3}}{3^{3/2}\sqrt{\pi}x^{7/3}} e^{-4/(27x^2)} U\left(\frac{1}{6},\frac{4}{3},\frac{4}{27x^2}\right).
\label{LevyU}
\end{equation}
From this, it is easy to get the Airy distribution since ${\cal L}^{-1}[s\tilde{F}(s)] = F'(x) + F(0)$.  Given that $L_{2/3,1}(0)=0$, we then have, using Eq. (\ref{Ptilde0}),
\begin{align}
P(A,t) &= 2\sqrt{\pi}\sum_k\left. \frac{1}{A_0 \lambda_k^{3}}\frac{d}{dx} L_{2/3,1} (x)\right|_{x=\hat{A}/\lambda_k^{3/2}}\nonumber\\
&= 2\sqrt{\pi}  \left(\frac{2^{4/3}}{3^{3/2}\sqrt{\pi} x^{7/3}}\right) \sum_k\frac{2}{\lambda_k^3 x} e^{-4/(27x^2)} \left.U\left(-\frac{5}{6},\frac{4}{3},\frac{4}{27x^2}\right)\right|_{x=\hat{A}/\lambda_k^{3/2}} \nonumber\\
&=   \left(\frac{2^{10/3}}{3^{3/2}\hat{A}^{10/3}A_0}\right) \sum_k \lambda_k^2  e^{-4\lambda_k^3/(27\hat{A}^2)} U\left(-\frac{5}{6},\frac{4}{3},\frac{4\lambda_k^3}{27\hat{A}^2}\right) .
\label{airy0}
\end{align}
Substituting $D=1/2$ into Eq.  (\ref{airy0}) reduces to the known result~\cite{Takacs}.

It is interesting and somewhat amusing to note that there is yet another alternative representation of $L_{2/3,1}$ in terms of the Airy function, $\textrm{Ai}(\cdot)$:
\begin{equation}
L_{2/3,1}(x)=6\zeta^{7/4}\left(\textrm{Ai}(\zeta)-\frac{\textrm{Ai}'(\zeta)}{\sqrt{\zeta}}\right)e^{-2\zeta^{3/2}/3}
\label{LevyAi}
\end{equation}
where $\zeta \equiv (3x)^{-4/3}$. To derive this, we start with Eq. (\ref{LevyU}) and apply the identity~\cite{AbS}
\begin{equation}
U(a,b,z) = z^{1-b}U(1+a-b,2-b,z)
\end{equation}
to transform $U(1/6,4/3,\cdot)$ to $U(-1/6,2/3,\cdot)$.
We then apply the identity~\cite{AbS}
\begin{equation}
U(a-1,b-1,z) = (1-b+z)U(a,b,z) - zU'(a,b,z)
\end{equation}
to transform $U(-1/6,2/3,\cdot)$ to $U(5/6,5/3,\cdot)$ and its derivative.  Lastly, we use the identity~\cite{AbS}
\begin{equation}
U\left(\frac{5}{6},\frac{5}{3},\frac{4}{3}z^{3/2}\right) = \frac{\sqrt{\pi} }{z} e^{2z^{3/2}/3} 2^{-2/3}3^{5/6} \textrm{Ai}(z)
\end{equation}
to convert the Kummer $U$ function and its derivative to the Airy function and its derivative.  An alternate demonstration of the equivalence of the two forms is to notice that $H(y)\equiv L_{2/3,1}(x(y))$, where $y(x)\equiv4/(27x^2)$, satisfies the differential equation
\begin{equation}
y\partial_y^2 H - (1-y)\partial_y H + \frac{35}{36y}H=0
\end{equation}
with the normalization fixed by considering the small $y$ behavior of the solution, and that our alternate form also satisfies the same equation with the same small $y$ behavior.

We can as before obtain $P(A,t)$ from $L_{2/3,1}(x)$ by  differentiating, this time using our alternate form Eq. (\ref{LevyAi}):
\begin{equation}
P(A,t)=\frac{12\sqrt{\pi}}{A_0} \sum_k \frac{\zeta_k^{5/2}}{\lambda_k^3}e^{-2\zeta_k^{3/2}/3}\left[(8\zeta_k^{3/2}-7)\textrm{Ai}(\zeta_k) - (8\zeta_k^{3/2}-5)\frac{\textrm{Ai}'(\zeta_k)}{\sqrt{\zeta_k}}\right]\Bigg|_{\zeta_k=\lambda_k^2(3\hat{A})^{-4/3}} .
\end{equation}
We find this result  poetic, since it expresses the Airy distribution directly in terms of the Airy function, which with it is intimately connected, as the $\lambda_k$ in this case are the absolute value of the zeros of the Airy function.

\section{Asymptotics of $d_k$, $\lambda_k$\label{sechik}}
Our formula for the Bessel distribution depends on the solution of the time-independent Schr\"odinger eqn., Eq. (\ref{Schro}) through the $\lambda_k$, the eigenvalues,
and the $d_k$, characterizing the small $x$ behavior of $\phi_k$ (see Eq. (\ref{dk})).  In this section, we study the asymptotic behavior of these quantities for large $k$.

\begin{figure}
\includegraphics[width=0.7\textwidth]{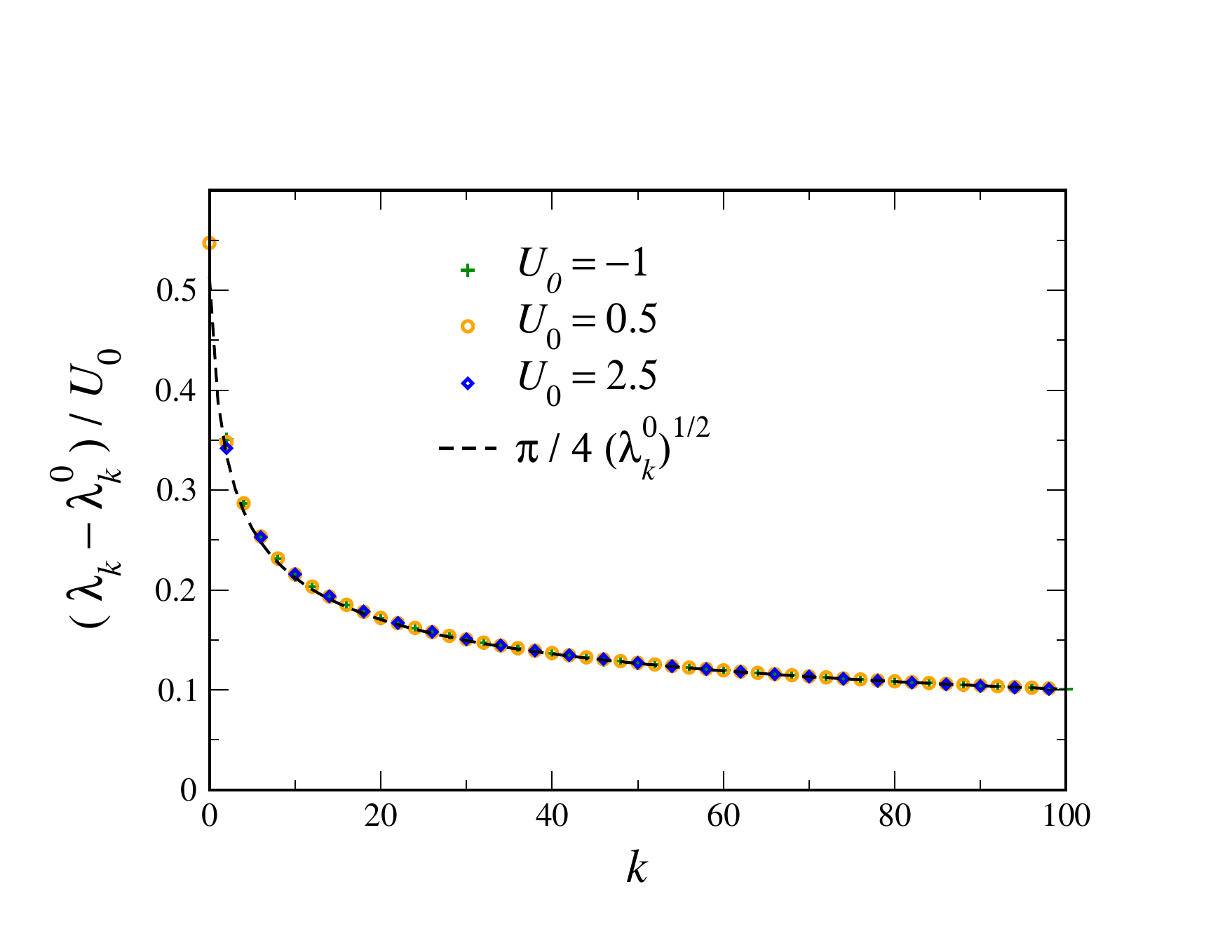}
\caption{The scaled energy difference, $(\lambda_k - \lambda_k^0)/U_0$ vs. $k$ for $U_0=-1$, $0.5$,  and $2.5$.  Also shown is the analytic prediction
$\pi/(4\sqrt{\lambda_k^0})$, where -$\lambda_k^0$, $k=0, 1 ,2 \ldots$ is the $k$th zero of the Airy function.}
\label{fig_lambda}
\end{figure}

Due to the singular nature of the effective potential of the Schr\"odinger eqn. near the origin, we cannot simply use the WKB approximation~\cite{Bender}, even for large $k$.  Rather, we solve the Schr\"odinger eqn. for small $x$, where the linear term in the potential is negligible, and for large $x$, where the $1/x^2$ term can be dropped, and match in the middle where $x$ is of order unity. Near the origin, we have
\begin{equation}
\phi_k \approx A_k \sqrt{x} J_{|\alpha|}(\lambda_k^{1/2} x) ,
\label{phiinner}
\end{equation}
This is valid as long as $x \ll \lambda_k$, since then the linear term in the potential is much smaller than the energy.  For large $x$, we have
\begin{equation}
\phi_k \approx B_k \textrm{Ai}(x-\lambda_k) .
\label{airy}
\end{equation}
This is valid as long as the $1/x^2$ term in the effective potential is much smaller than the energy, i.e., $x\gg \lambda_k^{-1/2}$.  
These two approximations must match in the overlap region, namely $\lambda_k^{-1/2} \ll x \ll \lambda_k$.  In this region, $ \lambda_k^{1/2} x \gg 1$ and the Bessel solution, Eq. (\ref{phiinner}) becomes~\cite{AbS}
\begin{equation}
\phi_k \approx A_k \sqrt{\frac{2}{\pi \lambda_k^{1/2}}} \cos\left(\lambda_k^{1/2} x - \frac{|\alpha| \pi}{2} - \frac{\pi}{4}\right) .
\end{equation}
Similarly, as $x\ll \lambda_k$, we are deep in the interior of the linear well, and  Eq. (\ref{airy})  becomes~\cite{AbS}
\begin{equation}
\phi_k \approx B_k \pi^{-1/2} \lambda_k^{-1/4}\sin\left(\frac{2}{3}\lambda_k^{3/2} - \lambda_k^{1/2} x + \frac{\pi}{4}\right) .
\end{equation}
Matching these two solutions for large $\lambda_k$ yields our first main result of the section:
\begin{equation}
\frac{2}{3}\lambda_k^{3/2} \approx \pi\left(k + \frac{|U_0+1| + 2}{4}\right) ,
\label{eqlk}
\end{equation}
so that $\lambda_k$ grows as $k^{2/3}$.  Setting $U_0=0$ gives a result that is of course consistent with the standard asymptotic formula~\cite{AbS} for the zeros of the Airy function, namely
\begin{equation}
\frac{2}{3}(\lambda_k^0)^{3/2} \approx \pi (k + 3/4) .
\end{equation}
Another way to express this is to calculate the difference  between $\lambda_k$ and its Airy $U_0=0$ value,  $\lambda_k^0$:
\begin{equation}
\lambda_k - \lambda_k^0 \approx \frac{\pi( |U_0+1|-1)}{4\sqrt{\lambda_k^0}} .
\label{delE}
\end{equation}
We test this formula in Fig. \ref{fig_lambda}, where we plot $(\lambda_k - \lambda_k^0)/U_0$ for various positive values of $U_0$, together with the prediction
$\pi/(4\sqrt{\lambda_k^0})$.
Comparing our result, Eq. (\ref{eqlk}),  to  the standard WKB answer, 
\begin{equation}
\int_{x_L}^{x_R}dx\, \sqrt{\lambda_k - \frac{U_0(U_0+2)}{4x^2} -x} = \pi (n+1/2)
\end{equation}
where $x_L$, $x_R$ are the two turning points of the effective potential,
and using the fact that for large 
$\lambda_k$, 
\begin{equation}
\int_{x_L}^{x_R}dx\, \sqrt{\lambda_k - \frac{U_0(U_0+2)}{4x^2} -x}  \approx \frac{2}{3}\lambda_k^{3/2} - \frac{\pi}{4}\sqrt{U_0(U_0+2)} ,
\end{equation}
we see that we get agreement only for large $|U_0|$, and there only to leading order in $|U_0|$.  This is a sign of the breakdown of the WKB ansatz that
$(\ln \phi_k)'' \ll [(\ln \phi_k)']^2$ in the small-$x$ regime, except when $|U_0| \gg 1$.
In addition, Eq. (\ref{delE}) recovers the first-order perturbative answer for small $|U_0|$, calculated via $\lambda_k-\lambda_k^0 \approx U_0(U_0+2)/4\int_0^\infty (\phi_k^0(x))^2/x^2 dx$.

\begin{figure}
\includegraphics[width=0.7\textwidth]{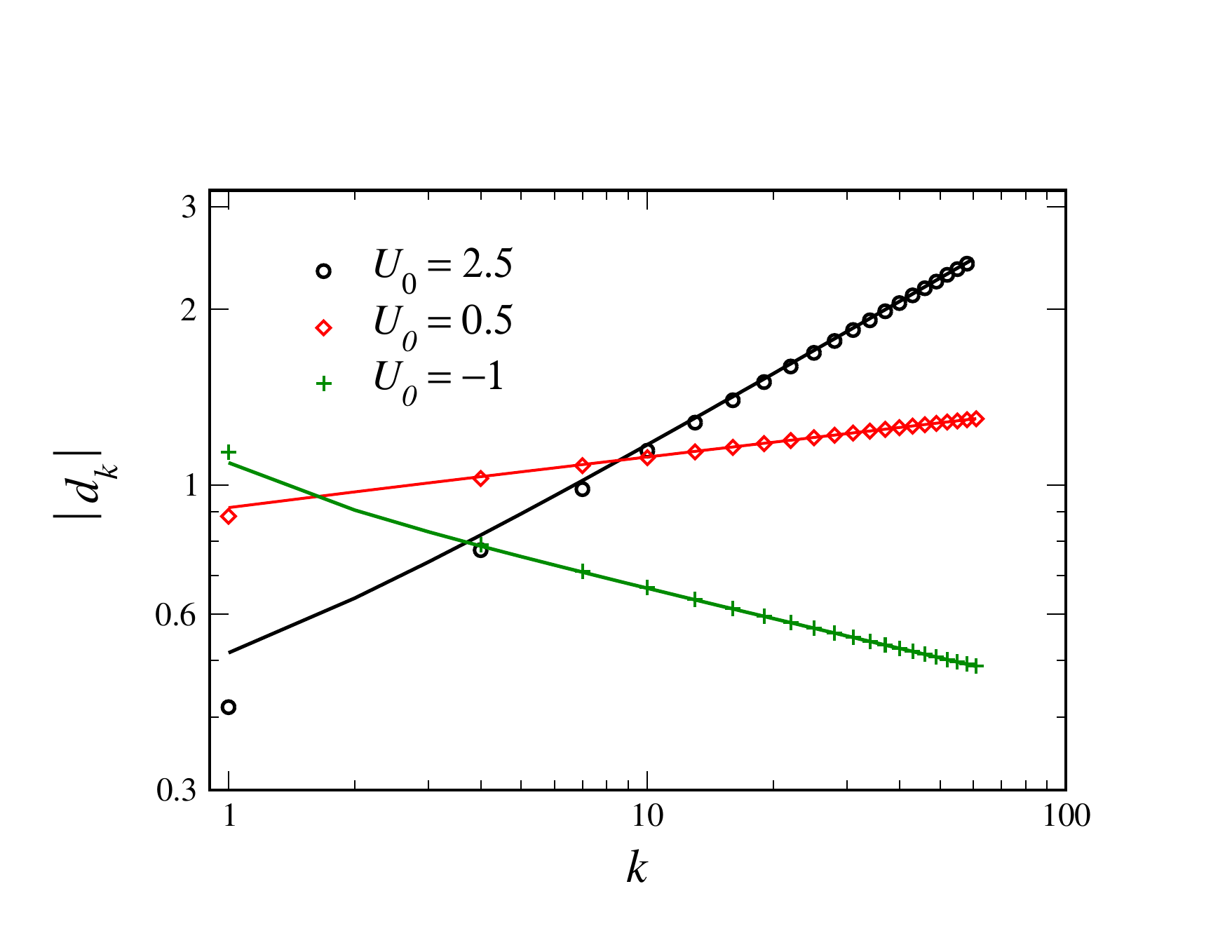}
\caption{Numerical results for $|d_k|$ vs. $k$ in log-log scale for the cases $U_0=-1$ (pluses), $0.5$ (diamonds),  and $2.5$ (circles), together with the analytic approximation, Eq. (\ref{eqdk}), using the numerically exact $\lambda_k$, shown as lines.}
\label{figdk}
\end{figure}

We now move on to calculate the asymptotics of $d_k$.
Proceeding with the matching, comparing the coefficients, we get
\begin{equation}
A_k = B_k 2^{-1/2} .
\end{equation}
The coefficient $B_k$ is determined by the normalization condition and is to leading order the same as for the Airy equation and so is given to this order by 
\begin{equation}
B_k \approx \frac{\sqrt{\pi}}{\lambda_k^{1/4}} .
\end{equation}
Using the small argument expansion of $J_\alpha$~\cite{AbS}:
\begin{equation}
J_{\alpha}(z) \approx \frac{1}{\Gamma(1+\alpha)}\left(\frac{z}{2}\right)^\alpha ,
\end{equation}
the small $x$ behavior of $\phi_k$ is given by
\begin{equation}
\phi_k \approx \sqrt{\frac{x}{2}}\cdot\frac{\sqrt{\pi}}{\lambda_k^{1/4}}\cdot \frac{1}{\Gamma(1+|\alpha|) }\left(\frac{\lambda_k^{1/2} x}{2}\right)^{|\alpha|}  = \left(\frac{x}{2}\right)^{|\alpha|+1/2} \frac{\sqrt{\pi}\lambda_k^{(2|\alpha|-1)/4}}{\Gamma(1+|\alpha|) },
\end{equation}
from which we read off
\begin{equation}
d_k \approx 2^{-|\alpha|-1/2}\frac{\sqrt{\pi}\lambda_k^{(2|\alpha|-1)/4}}{\Gamma(1+|\alpha|)} .
\label{eqdk}
\end{equation}
In Fig. \ref{figdk}, we present numerical results for $d_k$ for $U_0=-1$, $0.5$ and $2.5$, together with the analytic approximation Eq. (\ref{eqdk}).  The agreement is excellent, not only as expected for the largest $k$'s, but also for intermediate $k$'s, i.e. down to $k\sim2$ for $|U_0|\le 1$.

\section{Integer Moments of the Bessel Distribution}
We can use our results for the asymptotics of $d_k$ and $\lambda_k$ to verify the normalization of $P(A,t)$, using the Laplace transform representation.
Here we need to show that $\lim_{s \to 0} \tilde{P}(s,t) = 1$.  In the limit of small $s$,  the sum is dominated by the large $k$ terms. Thus, we can replace the sum over $k$ by an integral and use the large $k$ approximations for $\lambda_k$, Eq. (\ref{eqlk}) and $d_k$, Eq. (\ref{eqdk}):
\begin{align}
\lim_{s\to 0} \widetilde{P}(s,t) &\approx 2^{2|\alpha| + 1} \Gamma(|\alpha|+1) (A_0 s)^{|\nu|+2/3} \int_0^\infty d\lambda \frac{dk}{d\lambda} 
 d_k^2 e^{-(A_0 s)^{2/3} \lambda_k} \nonumber\\
 &= \frac{\pi (A_0 s)^{|\nu| + 2/3}}{\Gamma(1+|\alpha|)} \int_0^\infty d\lambda \frac{\sqrt{\lambda}}{\pi} \lambda^{|\alpha|-1/2}e^{-(A_0 s)^{2/3} \lambda} \nonumber\\
 &=1 .
 \end{align}
 In principle, one could use the higher-order corrections to $d_k$ and $\lambda_k$ to calculate the first and higher moments of the distribution.  We however, will adopt a different tack to calculate the first and second moments, namely via Feynman's propagator expansion.
 
\begin{figure}
 \includegraphics[width=0.7\textwidth]{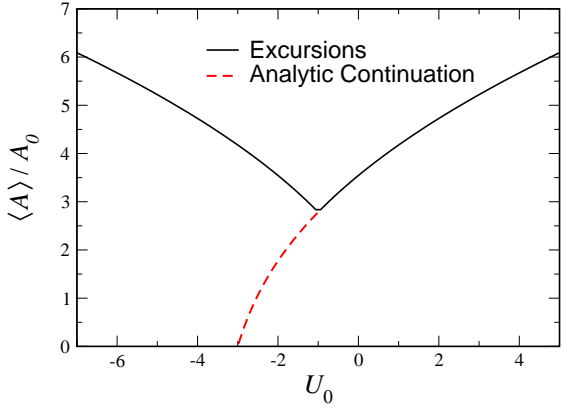}
 \caption{The scaled first moment of the area, ${\cal{M}}_1/A_0$ vs. $U_0$, from Eq. (\ref{eqM1}).  Also shown in the analytic continuation for $-3<U_0<-1$.}
 \label{figM1}
 \end{figure}

 The key point is that the $n$th moment of $P(A,t)$ is given (up to a sign) by the $n$th derivative of the Laplace transform $\widetilde{P}(s,T)$ with respect to $s$, evaluated at $s=0$.  Since $\widetilde{P}(s,T)$ is given in terms of the propagator $K_s(x,x_0;T)$,  if we expand the latter in powers of $s$, we have what we need.  This is just the perturbative expansion of the
 propagator in the linear potential $sx$. The  ``unperturbed" propagator is just $K_0(x,x';t)$, given in Eq. (\ref{K0}).  Then, to linear order in $s$,
 \begin{equation}
 K(x,x_0;T) \approx K_0(x,x_0;T) - s \int_0^\infty dx_1\int_0^T dt_1  K_0(x,x_1;T- t_1) x_1 K_0(x_1,x_0;t_1) 
 \end{equation}
 and formally, in operator notation,
 \begin{equation}
 K = K_0 - s K_0 x K_0 + s^2 K_0 x K_0 x K_0 - s^3 K_0 x K_0 x K_0 x K_0 + \ldots
 \end{equation}
 generates the entire expansion in $s$.  The moments are then given by the $x=x_0\to 0$ limits of the terms of this expansion, suitably normalized by the same limit of $K_0$ (see Eqs. (\ref{Ptilde}), (\ref{Gs}) and (\ref{G0})). Thus, using the small argument expansion of $I_\alpha(z) \sim (z/2)^\alpha/\Gamma(1+\alpha)$, and rescaling times by $T$ and
 lengths by $\sqrt{DT}$, the first moment is given by
 \begin{align}
 {\cal{M}}_1 &= \lim_{x=x_0\to 0} \frac{\int_0^\infty dx_1\int_0^T dt_1  K_0(x,x_1;T-t_1) x_1 K_0(x_1,x_0;t_1)}{K_0(x,x_0)} \nonumber\\
 &= \frac{A_0}{2^{2|\alpha|+1}\Gamma(1+|\alpha|)} \int_0^\infty dx_1 \int_0^1 dt_1 \frac{x_1^{2+2|\alpha|}}{[(1-t_1)t_1]^{|\alpha|+1}} \exp\left(- \frac{x_1^2}{4(1-t_1)} - \frac{x_1^2}{4t_1}\right) \nonumber\\
 &= \frac{A_0}{2^{2|\alpha|+1}\Gamma(1+|\alpha|)} \int_0^1 dt_1 \frac{1}{2} \left[4t_1(1-t_1)\right]^{|\alpha|+3/2} \Gamma\left(|\alpha| + \frac{3}{2}\right) \left[t_1(1-t_1)\right]^{-|\alpha| - 1} \nonumber\\
 &= \frac{2A_0 \Gamma(|\alpha|+3/2)}{\Gamma(|\alpha|+1)} \int_0^1 dt_1 \sqrt{t_1(1-t_1)} \nonumber \\
 &= \frac{\pi\Gamma(|\alpha|+3/2)}{4\Gamma(|\alpha| + 1)} A_0 .
 \label{eqM1}
 \end{align}
 This reduces to the known answer for the first moment of the Airy Distribution when $\alpha=1/2$.    For large $|\alpha|$, ${\cal{M}}_1$ grows as $\sqrt{|\alpha|}$, i.e. $\sqrt{|U_0|}$. A graph is presented in Fig. \ref{figM1}.
 
\begin{figure}
 \includegraphics[width=0.7\textwidth]{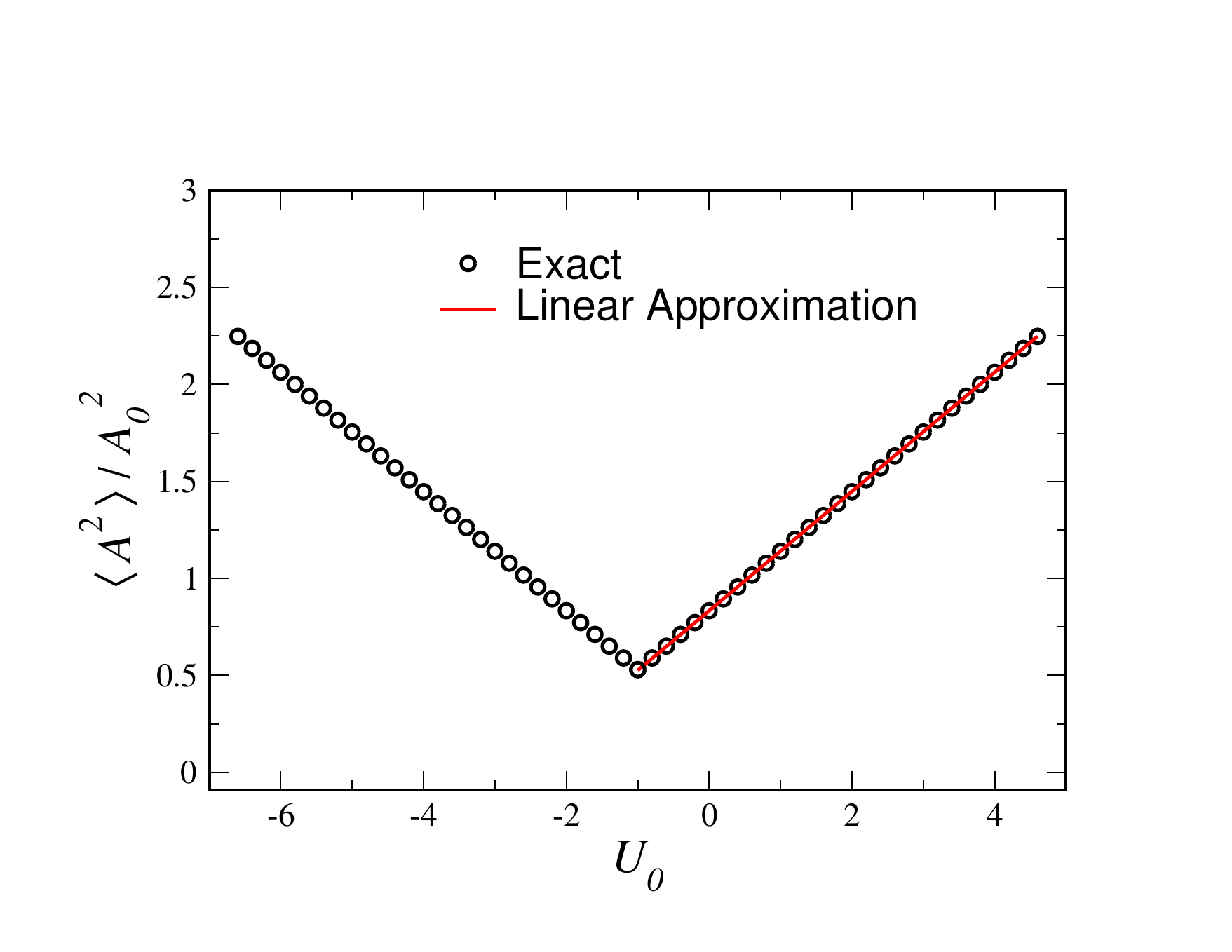}
 \caption{${\cal{M}}_2/A_0^2$ vs. $U_0$, via a numerical calculation of Eq. (\ref{eqM2}).  Also shown in the linear approximation, Eq. (\ref{eqM2lin}).}
 \label{figM2}
 \end{figure}

 The calculation of the second moment is similar, but much more involved algebraically.
 \begin{align}
 {\cal{M}}_2 &= \frac{A_0^2}{2^{2|\alpha|+1}\Gamma(1+|\alpha|)} \int_0^\infty dx_2 \int_0^1 dt_2 \int_0^\infty dx_1 \int_0^{1-t_2}  dt_1\, I_{|\alpha|}\left(\frac{x_2 x_1}{2t_2}\right) \times\nonumber\\
& \qquad \frac{x_2^{|\alpha| + 2}x_1^{|\alpha|+2}}{\left[t_1  (1-t_1-t_2)\right]^{|\alpha|} t_1 t_2 (1-t_1-t_2)}
 \exp\left(-\frac{x_2^2(1-t_1)}{4(1-t_1-t_2)} - \frac{x_1^2(t_2+t_1)}{4t_1t_2}\right). \nonumber\\
 \end{align}
 To proceed, we expand the Bessel function in a Froebenius series, and integrate term by term over $x_1$ and $x_2$, giving
 \begin{align}
 {\cal{M}}_2 &= \frac{A_0^2}{2^{2|\alpha|+1}\Gamma(1+|\alpha|)} \int_0^1 \!dt_2 \int_0^{1-t_2} \!dt_1\sum_{k=0}^\infty \left[\frac{ (1-t_1-t_2)(t_2)^2t_1}{(1-t_1)(t_1+t_2)}\right]^{|\alpha|+k+3/2}\times\nonumber\\ 
 &\qquad\qquad\qquad\qquad\qquad\qquad\frac{\left[\frac{1}{2}\Gamma(|\alpha|+k+3/2)4^{|\alpha|+k+3/2}\right]^2}{4^{|\alpha|+2k} k!\Gamma(|\alpha|+k+1) t_1^{|\alpha|+1}t_2^{|\alpha|+2k+1}(1-t_1-t_2)^{|\alpha|+1}} \nonumber\\
 &= \frac{8A_0^2}{\Gamma(1+|\alpha|)} \sum_{k=0}^{\infty} \Gamma^2(|\alpha|+k+3/2) \times\nonumber\\
 & \qquad\qquad\qquad\qquad\int_0^1 dt_2 \int_0^{1-t_2} dt_1 \frac{t_1^{k+1/2} t_2^{|\alpha|+2} 
 (1-t_1-t_2)^{k+1/2}}{k!\Gamma(|\alpha|+k+1) [(1-t_1)(t_1+t_2)]^{|\alpha|+k+3/2}}\nonumber\\
 &= \frac{8A_0^2}{\Gamma(1+|\alpha|)} \sum_{k=0}^\infty 2^{2k} \Gamma^2(|\alpha|+k+3/2) \times\nonumber\\
 &\qquad\qquad\qquad\qquad\int_0^1 dt_3 \int_0^{1-t_3} dt_1 \frac{t_1^{k+1/2} (1-t_1-t_3)^{|\alpha|+2} 
 t_3^{k+1/2}}{k!\Gamma(|\alpha|+k+1) [(1-t_1)(1-t_3)]^{|\alpha|+k+3/2}} ,
 \end{align}
 where we have changed variables to $t_3=1-t_2-t_1$.  Integrating now over $t_1$ yields a hypergeometric function and the subsequent integration over
 $t_3$ yields a generalized hypergeometric function:
 \begin{align}
 {\cal{M}}_2 &=  \frac{16\Gamma(3+|\alpha|)A_0^2}{\Gamma(1+|\alpha|)} \sum_k \frac{\Gamma^2(k+3/2) \Gamma^2(|\alpha|+k+3/2)}{k!\Gamma(k+9/2)\Gamma(|\alpha|+k+1)\Gamma(|\alpha|+k+9/2)}\times\nonumber\\
 &\qquad\qquad\qquad\qquad\qquad {}_3F_2\left(k+\frac{3}{2},|\alpha| + k + \frac{3}{2},3;k+\frac{9}{2},|\alpha|+k+\frac{9}{2};1\right).
 \label{eqM2}
 \end{align}
 The asymptotic behavior of the summand is $1/60k^2 - (9+4|\alpha|)/240k^3$, which can be obtained from the large-$k$ behavior of the hypergeometric function
 \begin{equation}
 {}_3F_2\left(k+\frac{3}{2},|\alpha| + k + \frac{3}{2},3;k+\frac{9}{2},|\alpha|+k+\frac{9}{2};1\right) \approx \frac{k^3}{60} +\frac{k^2}{40}(8 + |\alpha|) .
 \end{equation}
 With this asymptotic formula for the high-$k$ terms, which allows the summand to be approximated for large $k$, the sum can be easily performed by breaking the sum at some large $K$, summing the low-$k$ terms numerically and using this large-$k$  formula to perform the second sum analytically. We  calculate ${\cal{M}}_2$ as a function of $|\alpha|$ in this manner, and the results are displayed in Fig. \ref{figM2}.
 
 A numerical evaluation shows that for $U_0=0$, i.e., $\alpha=1/2$, ${\cal{M}}_2$ is consistent with the known Airy answer of $5A_0^2/6$~\cite{Takacs}.  We will present an alternate derivation of this result from our formalism later. There is one other value of $\alpha$ for which we can derive an exact answer, namely
 $\alpha=3$. For this case, the upper index, $\alpha + k + 3/2$ of the generalized hypergeometric function in Eq. (\ref{eqM2}) equals the lower index $k + 9/2$, and the $(3,2)$ hypergeometric function reduces to a standard $(2,1)$ hypergeometric function. Then
 \begin{align}
 {\cal{M}}_2 &= \frac{16\Gamma(6)A_0^2}{\Gamma(4)} \sum_k \frac{\Gamma^2(k+3/2) \Gamma^2(k + 9/2)}{k!\Gamma(k+9/2)\Gamma(k+4)\Gamma(k+15/2)} {}_2F_1\left(k+\frac{3}{2},3;k+\frac{15}{2};1\right) \nonumber\\
 &= 320A_0^2 \sum_k \frac{\Gamma^2(k+3/2)\Gamma(k+9/2)}{k!\Gamma(k+4)\Gamma(k+15/2)}\cdot\frac{\Gamma(k+15/2)\Gamma(3)}{\Gamma(6)\Gamma(k+9/2)}\nonumber\\
 &= \frac{16}{3} A_0^2 \sum_k \frac{\Gamma^2(k+3/2)}{k!\Gamma(k+4)} \nonumber\\
 &= \frac{16}{3} A_0^2 \frac{\Gamma^2(3/2)}{\Gamma(4)} {}_2F_1(3/2,3/2;4;1) \nonumber\\
 &=\frac{16}{3} A_0^2 \frac{\Gamma^2(3/2)}{\Gamma(4)}\cdot \frac{\Gamma(4)\Gamma(1)}{\Gamma^2(5/2)} \nonumber\\
 &= \frac{64}{27}A_0^2 \ .
 \end{align}
 
  For large $|\alpha|$, the sum is dominated by $k \sim {\cal{O}}(|\alpha|)$.  Evaluating the summand in this limit yields
 \begin{align}
 {\cal{M}}_2 &  \approx 8A_0^2\sum_k \sqrt{k(k+|\alpha|)}\left[(6k^2 + 6|\alpha| k + |\alpha|^2)\ln \frac{k + |\alpha|}{k} - |\alpha|(6k + 3|\alpha|)\right]\nonumber\\
 &\approx 8A_0^2 |\alpha| \int_0^\infty dx \sqrt{x(x+1)}\left[(6x^2 + 6 x + 1)\ln \frac{x + 1}{x} - |\alpha|(6x + 3)\right] \nonumber\\
 &= 0.61685 |\alpha| A_0^2\ .
 \end{align}
 There is also a constant contribution, which is much more difficult to evaluate.  What is striking is how linear ${\cal{M}}_2(|\alpha|)$ is.  One measure of this is the fact that the slope of the line connecting the exact results for $|\alpha|=1/2$ and $|\alpha|=3$ is $(64/27 - 5/6)  A_0^2 / (5/2) = (83/135)  A_0^2 \approx 0.6148 A_0^2$, so that
\begin{equation}
{\cal{M}}_2 \approx   A_0^2 \left[\frac{83}{135}\left(|\alpha|-\frac{1}{2}\right) + \frac{5}{6}\right] .
\label{eqM2lin}
\end{equation}
In Fig. \ref{figM2} we see that this works excellently.

\section{The Analytically Continued Bessel Distribution\label{secU0m3}}
As our results for the moments make clear, the Bessel Distribution as we have defined it is non analytic at $U_0=-1$, corresponding to $d=2$.  This is because the boundary condition, Eq. (\ref{eqbc}), is nonanalytic in $U_0$. One can consider an alternate distribution for $U_0<-1$, defined by the analytic
continuation of the results for $U_0>-1$.  This corresponds to the requirement that the small-$x$ expansion of $\phi_k$ contain no $x^{-U_0/2}$ mode, which for $U_0<-1$ is the less singular mode.  In this case, we eliminate all the absolute values in all our analytic expressions above.  The analytically continued first moment is presented in Fig. \ref{figM1}.  In particular, we see
that the first  moment of this analytically continued distribution, which we shall label as $Q(A)$, vanishes in the limit $U_0\to -3$, corresponding to $d=4$. 
This implies that the $Q$ distribution must approach a $\delta$-function at $A=0$ in the limit, since the distribution is only defined for $A>0$. Thus, this limit is intriguing and in this section we examine it in more detail.

 The key point of the analysis is that, as we shall see, $\lambda_0 \sim {\cal{O}}(U_0+3)$, $d_0 \sim {\cal{O}}((U_0+3)^{1/2})$, $d_{k > 0} \sim {\cal{O}}(U_0+3)$, and $\lambda_{k>0}
\sim {\cal{O}}(1)$.  The behavior of $\lambda_{k>0}$ and $d_{k>0}$ are consistent with the large-$k$ asymptotics we worked out above.  Not unexpectedly,
the $k=0$ values qualitatively differ from the large-$k$ results, though the asymptotics do predict correctly that both $\lambda_0$ and $d_0$ vanish in the limit.

Assuming in accord with the results of numerical calculations that indeed $\lambda_0$ vanishes, we can construct the solution $\phi_0$ to the Schr\"odinger Eq.
(\ref{Schro}) to first order in $\lambda_0$, 
\begin{equation}
\phi_0 \approx \phi_0^0 + \lambda_0 \phi_0^1 .
\end{equation}
  To zeroth order, we have the zero-energy solution
\begin{equation}
\phi_0^0 = C \sqrt{x} K_{|\nu|}\left(\frac{2}{3}x^{3/2}\right)\ .
\end{equation}
This zero-energy solution is not an eigenfunction because of the term $x^{1/2+|\nu|}=x^{-U_0/2}$ (since $\nu<0$ for the $U_0$'s of interest here) in the small-$x$ expansion, which violates the boundary condition at small $x$.  This terms needs to be cancelled by the first-order contribution to the solution, induced by the small $\lambda_0$.  The correction $\phi_0^1$ satisfies the inhomogeneous equation
\begin{equation}
-{\phi_0^1}'' + \frac{U_0(U_0+2)}{4x^2} \phi_0^1 + x \phi_0^1 =  \phi_0^0\ ,
\end{equation}
with the solution
\begin{align}
\phi_0^1 = -\frac{2}{3}C \sqrt{x} &\Bigg[ -I_{|\nu|}\left(\frac{2}{3}x^{3/2}\right) \int_x^\infty dx' x'\left(K_{|\nu|}\left(\frac{2}{3}x'^{3/2}\right)\right)^2 
\nonumber\\
&\qquad {} + K_{|\nu|}\left(\frac{2}{3}x^{3/2}\right) \int_0^x dx' x' K_{|\nu|}\left(\frac{2}{3}x'^{3/2}\right) I_{|\nu|}\left(\frac{2}{3}x'^{3/2}\right)\Bigg]\ ,
\label{phi11}
\end{align}
where the factor $-2/3$ comes from the Wronskian of the two homogeneous solutions.  The key $x^{-U_0/2}$ term comes from the first of the
two terms in Eq. (\ref{phi11}).  Using $K_\nu(x) = (\pi/2)(I_{-\nu}(x) - I_{\nu}(x))/\sin(\pi\nu)$, we get the following equation for $\lambda_0$:
\begin{equation}
0 = -\frac{\pi}{2\sin(2\pi/3)} + \lambda_0\frac{2}{3} \int_0^\infty dx' x' \left(K_{|\nu|}\left(\frac{2}{3}x'^{3/2}\right)\right)^2 \approx -\frac{\pi}{\sqrt{3}} + \frac{2}{3}\lambda_0 \frac{3^{4/3}\Gamma^2(2/3)}{4 (U_0+3)}\  ,
\end{equation}
giving
\begin{equation}
\lambda_0 \approx \frac{2\pi}{3^{5/6}\Gamma^2(2/3)}(U_0+3) \ .
\end{equation}
The value of $d_0$ comes from the leading-order term.  Normalization gives
\begin{equation}
C = \left[ \int_0^\infty \left[x K_{|\nu|}^2\left(\frac{2}{3}x^{3/2}\right)\right]\right]^{-1/2} \approx \left[\frac{\Gamma^2(2/3) 3^{4/3}}{4(U_0+3)}\right]^{-1/2} 
= \frac{2}{\Gamma(2/3)3^{2/3}}\sqrt{U_0+3}\ ,
\end{equation}
and therefore
\begin{equation}
d_0 \approx C \frac{\Gamma(2/3) 3^{2/3}}{2} = \sqrt{U_0+3} \ .
\end{equation}

To investigate the $U_0\to-3$ behavior of the other $\lambda_k$, $d_k$, it is instructive to examine the first few terms of the Froebenius expansion about $x=0$ of $\phi_k$. Using the fact that there is no $x^{-U_0/2}$ term in the expansion, we find from Eq. (\ref{Schro}):
\begin{equation}
\phi_k = d_k x^{U_0/2} \left[ x - \frac{\lambda_k}{2(U_0+3)}x^3 + \frac{1}{3(U_0+4)}x^4 + \frac{\lambda_k^2}{8(U_0+3)(U_0+5)}x^5 + \ldots\right]\ ,
\end{equation}
where we have imposed the condition that there is no $x^{-U_0/2}$ term in the expansion.  We see that there are terms in $\phi_k$ which are of order
$1/(U_0+3)$ and there are terms of order 1.  We assume here that $\lambda_k$ remains finite in the limit $U_0\to -3$; otherwise the problem reduces to that treated above, yielding the unique value $\lambda_0$. Thus the $1/(U_0+3)$ terms dominate as $U_0 \to -3$, and these terms satisfy the equation
\begin{equation}
-{\phi_k^0}'' + \frac{3}{4x^2} \phi_k^0 + x\phi_k^0 = \lambda_k \phi_k^0 \ ,
\end{equation}
with the boundary condition that $\phi_k^0 \sim x^{3/2}$ as $x \to 0$ (and not as $x^{1+U_0/2}$ as does the exact solution!).  This Schr\"odinger equation defines an eigenvalue problem, which yields the leading order values of the $\lambda_k$ for $k=1,2,\ldots$.  The lowest eigenvalue is $\lambda_1 \approx 2.887$, so all the eigenvalues are positive.  Thus we
indeed confirm that all the $\lambda_{k > 0}$ have finite limits as $U_0\to -3$.  The $x^{1+U_0/2}$ term in the expansion is down by a factor of $U_0+3$ relative to $\phi_k^0$, and so $d_{k>0}$ is vanishes as $U_0+3$ in the limit.

With these results in hand, we can now turn to the leading order approximation to the Bessel Distribution.  In the Laplace representation, Eq. (\ref{Ptilde1}),
the prefactor $\Gamma(1+|\alpha|)$ diverges linearly in the limit, but $d_k^2$ vanishes quadratically for $k>0$ so all these terms do not contribute to leading order.  Thus, in the limit, we have
\begin{equation}
\widetilde{P}(s,T) \approx e^{-\lambda_0 \hat{s}^{2/3}} .
\end{equation}
We recognize this as the Laplace transform of the one-sided L\'evy $\alpha$-stable distribution with index $2/3$ and scale factor $A_0 \lambda_0^{3/2}$, i.e.,  of order $A_0 (U_0+3)^{3/2}$.  The integer moments of this distribution
all diverge.  This is because these moments are all dominated by the tail behavior, which is not described by the leading-order distribution we have calculated.  In truth, the full distribution for all $U_0<3$ possesses a Gaussian tail at $A/A_0$ of order 1.  The total weight of this tail is of order $U_0+3$,
so all the integer moments vanish as $U_0+3$, as we have seen explicitly in the case of the first and second moments.  The appearance of the limiting fat-tailed L\'evy $\alpha$-stable distribution is a result of the critical nature of the return properties of the $d=4$ Brownian random walk.
 
 \section{Integer Moments of the Airy Distribution}
 Our formalism allows for a fairly straightforward alternate calculation of the known integer moments of the Airy Distribution, using a Fourier representation of the $\delta$-function constraining the total time. For example, for the first moment,
 \begin{align}
 {\cal{M}}_1 &= \frac{A_0}{2\sqrt{\pi}} \int_0^\infty dt_1 \int_0^\infty dt_2 \int_0^\infty dx_1 \frac{x_1^3}{(t_1t_2)^{3/2}} e^{-x_1^2/4t_1-x_1^2/4t_2} \delta(t_1+t_2-1) \nonumber\\
&= \frac{A_0}{2\sqrt{\pi}} \int_0^\infty dt_1 \int_0^\infty dt_2 \int_0^\infty dx_1 \int_{-\infty-i\epsilon}^{\infty-i\epsilon} \frac{ds}{2\pi}e^{-is(t_1+t_2-1)} \frac{x_1^3}{(t_1t_2)^{3/2}} e^{-x_1^2/4t_1} e^{-x_1^2/4t_2} \nonumber\\
&= \frac{A_0}{4\sqrt{\pi^3}} \int_{-\infty-i\epsilon}^{\infty-i\epsilon} ds\, e^{is}  \int_0^\infty dx_1\, x_1^3 \left[2\frac{\sqrt{\pi} e^{-\sqrt{is} x_1}}{x_1}\right]^2\nonumber\\
&= \frac{A_0}{\sqrt{\pi}} \int_{-\infty-i\epsilon}^{\infty-i\epsilon} e^{is} \frac{1}{(2\sqrt{is})^2}\nonumber\\
&= \frac{A_0\sqrt{\pi}}{2} .
\end{align}
Similarly, for the second moment, using the fact that $I_{1/2}(x) = \sqrt{2/\pi x} \sinh x$,
\begin{align}
{\cal{M}}_2 &= \frac{A_0^2}{\pi}\hspace*{-0.1in} \prod_{i,j=\{1,2\}}  \int_0^\infty dt_i dx_j  \frac{x_1^2 x_2^2}{\sqrt{t_1^3 t_2 t_3^3}} e^{-\frac{x_1^2}{4t_1}-\frac{x_1^2}{4t_2} -\frac{x_2^2}{4t_2} - \frac{x_2^2}{4t_3}} \sinh\left(\frac{x_1x_2}{2t_2}\right) \delta(t_1+t_2+t_3-1) \nonumber\\
&= \frac{A_0^2}{4\pi^2}\prod_{i=\{1,2\}} \int_0^\infty dx_i \int_{-\infty - i\epsilon}^{\infty - i\epsilon}ds\, \left[ \left(\frac{2\sqrt{\pi}e^{-x_1\sqrt{is}}}{x_1}\right)\left(\frac{\sqrt{\pi}e^{-|x_1-x_2|\sqrt{is}}}{\sqrt{is}}\right)\left(\frac{2\sqrt{\pi}e^{-x_2\sqrt{is}}}{x_2}\right)  \right.\nonumber\\
&\ \qquad\qquad\qquad\qquad {} - \left. \left(\frac{2\sqrt{\pi}e^{-x_1\sqrt{is}}}{x_1}\right)\left(\frac{\sqrt{\pi}e^{-|x_1+x_2|\sqrt{is}}}{\sqrt{is}}\right)\left(\frac{2\sqrt{\pi}e^{-x_2\sqrt{is}}}{x_2}\right) \right] x_1^2 x_2^2 e^{is} \nonumber\\
&= \frac{A_0^2}{\sqrt{\pi}} \int_{-\infty-i\epsilon}^{\infty-i\epsilon} ds\, e^{is} \frac{1}{(is)^{5/2}} \left[\frac{3}{8} - \frac{1}{16}\right] \nonumber\\
&= \frac{5}{6}A_0^2 ,
\end{align}
in accord with the known results~\cite{Takacs}.
The higher integer moments of the Airy Distribution can easily be calculated along the same lines.

\section{The $\nu$th Moment of the Bessel Distribution}
Another moment of the Bessel Distribution can be calculated in closed form, namely the $\nu$th moment, where we restrict ourselves here to $\nu\ge 0$.  This is a generalization of the $1/3$ moment of the Airy Distribution, which has been calculated previously~\cite{Crandall,Flajolet}.  Our method is similar in spirit to this calculation, but different in detail.  Rather than employing an analytic continuation to deal with the singular nature of the calculation, we employ an explicit cutoff.  

Going back to our Laplace space representation of the  probability density, $\widetilde{P}(s)$, we break the sum into two pieces at $k=N\gg 1$, writing
$\widetilde{P}(s) = \widetilde{P}_N(s)+ \widetilde{P}_R(s)$.  The latter piece can be computed explicitly, using the large $k$ expressions for $d_k$, Eq. (\ref{eqdk}) and $\lambda_k$, Eq. (\ref{eqlk}):
\begin{align}
\widetilde{P}_R(s) &\approx    2^{2\alpha + 1} \Gamma(\alpha+1) (sA_0)^{\nu+2/3} \int_{\Lambda}^\infty \frac{\sqrt{\lambda} d\lambda}{\pi} \frac{2^{-2\alpha-1}\pi \lambda^{\alpha-1/2}}{\Gamma^2(\alpha+1)} e^{-\lambda (sA_0)^{2/3}}\nonumber\\
&= \frac{\Gamma(\alpha+1,(sA_0)^{2/3}\Lambda)}{\Gamma(\alpha+1)}\ ,
\end{align}
where $\Lambda$ is the eigenvalue of the $N+1$st mode, $\Lambda \approx (3\pi N/2)^{2/3}$.
The contribution to ${\cal{M}}_\nu$ from these large $k$ terms is then
\begin{align}
{\cal{M}}_R &= \int_0^\infty A^\nu P_R(A) dA = \int_0^\infty dA\, \frac{1}{\Gamma(1-\nu)}\int_0^\infty ds s^{-\nu} e^{-sA} AP_R(A) \nonumber\\
&= -\frac{1}{\Gamma(1-\nu)} \int_0^\infty ds s^{-\nu} \frac{d}{ds} \widetilde{P}_R(s) \nonumber\\
&= -\frac{1}{\Gamma(1-\nu)} \int_0^\infty dt t^{-\alpha} \frac{d}{dt} \frac{\Gamma(\alpha+1, A_0^{2/3} \Lambda t)}{\Gamma(\alpha+1)} \nonumber\\
&=  \frac{A_0^{\nu+2/3}\Lambda^{\nu+1}}{\Gamma(1-\nu)\Gamma(\alpha+1)} \int_0^\infty dt e^{-A_0^{2/3}\Lambda t} \nonumber\\
&=  \frac{(A_0\Lambda)^{\nu}}{\Gamma(1-\nu)\Gamma(\alpha+1)} \ .
\label{MR}
\end{align}
The singular growth of ${\cal{M}}_R$ with $\Lambda$ is the reason behind the need for the cutoff treatment.  We now need to treat the rest of the terms.
We assume $x<x'$ and write
\begin{equation}
\widetilde{P}_N(s) \equiv  2^{2\alpha + 1} \Gamma(\alpha+1) (sA_0)^{\nu+2/3} \lim_{x,x' \to 0} (x x')^{-U_0/2-1} \sum_{k=0}^N \phi_k(x)\phi_k(x') e^{-(sA_0)^{2/3} \lambda_k }\ .
\end{equation}
Since this is a finite sum, there is no problem exchanging orders of summation and integration in computing the contribution of these terms to the moment, and performing the integration as in Eq. (\ref{MR}), we find
\begin{align}
{\cal{M}}_N &= -\frac{2^{2\alpha + 1} \Gamma(\alpha+1) A_0^{\nu+2/3}}{\Gamma(1-\nu)}  \lim_{x,x'\to 0}  (xx')^{-U_0/2-1} \int_0^\infty dt\, t^{-\alpha} \times\nonumber\\
&\qquad\qquad\qquad\qquad\qquad\qquad\qquad\frac{d}{dt} \left[ t^{\alpha+1} \sum_{k=0}^N \phi_k(x)\phi_k(x') e^{-A_0^{2/3} \lambda_k t}\right] \nonumber\\
&= -\frac{2^{2\alpha + 1} \alpha\Gamma(\alpha+1)A_0^{\nu}}{\Gamma(1-\nu)} \lim_{x,x'\to 0}  \left[(xx')^{-U_0/2-1}  \sum_{k=0}^N \frac{\phi_k(x)\phi_k(x')}{\lambda_k}\right]\ .
\end{align}
The sum, if not cut off at $N$, would be simply the value at $E=0$ of the energy Green's function, $\widetilde{K}(x,x';E)$, i.e. the Laplace transform of the propagator $K$ (i.e., $K_s$, from Eq. (\ref{eqKs}) after rescaling out $s$ and $D$) with respect to the time:
\begin{equation}
\widetilde{K}(x,x';E) \equiv \int_0^\infty dt K(x,x';t) e^{-Et} = \sum_k \frac{\phi_k(x)\phi_k(x')}{E+\lambda_k}\ ,
\end{equation}
which satisfies the equation
\begin{equation}
\hat{H}\widetilde{K} + E\widetilde{K} = \delta(x-x')\ .
\end{equation}
The $E=0$ Green's function is given by
\begin{equation}
\widetilde{K}(x,x';0) = \frac{2}{3}\sqrt{xx'} I_\nu\left(\frac{2}{3} x_<^{3/2}\right) K_\nu\left(\frac{2}{3} x_>^{3/2}\right)\ ,
\end{equation}
where $I_\nu$ and $K_\nu$ are the modified Bessel functions, and $x_>$ ($x_<$) is the greater (lesser) of $x$ and $x'$.  For small $x$,$x'$, we can use
the standard Froebenius expansions of the Bessel functions to get, for $x'>x$:
\begin{equation}
\widetilde{K}(x,x';0) \approx \frac{1}{3\nu}\sqrt{x x'}\left(\frac{x}{x'}\right)^\alpha\left(1 - \left(\frac{x'^{3/2}}{3}\right)^{2\nu}\frac{\Gamma(1-\nu)}{\Gamma(1+\nu)}\right) \ .
 \end{equation}
 Calling the sum over the first $N$ terms $\widetilde{K}_N$, we can express this as
  as the difference between the zero-energy Green's function and the remainder piece, $\widetilde{K}_R$, i.e., the sum over the $k>N$ terms.
 \begin{equation}
 \widetilde{K}_R(x,x') \approx \int_\Lambda^\infty \frac{\sqrt{\lambda} d\lambda}{\pi} \frac{\phi_k(x) \phi_k(x')}{\lambda} \ .
 \end{equation}
 Substituting the large-$k$ form of the eigenfunctions, Eq. (\ref{phiinner}), we have
 \begin{align}
\widetilde{K}_R(x,x') &\approx \int_{\sqrt{\Lambda}}^\infty \frac{d\ell}{\ell} \sqrt{xx'} J_\alpha(\ell x) J_\alpha(\ell x')\nonumber\\
 &= \frac{x^{\alpha+1/2}}{2\alpha x'^{\alpha-1/2}} - \int_0^{\sqrt{\Lambda}} d\ell \sqrt{x x'}\frac{(\ell^2 x x'/4)^\alpha}{\ell \Gamma^2(\alpha+1)} \nonumber\\
 &= \frac{x^{\alpha+1/2}}{2\alpha x'^{\alpha-1/2}}  - (xx')^{\alpha+1/2} \frac{\Lambda^\alpha}{2^{2\alpha+1} \alpha \Gamma^2(\alpha+1)} \ .
 \end{align}
As $\widetilde{K}=\widetilde{K}_N+\widetilde{K}_R$, we have
\begin{align}
\widetilde{K}_N(x,x') &= \widetilde{K}(x,x';0) - \widetilde{K}_R(x,x') \nonumber\\
&= -\frac{\Gamma(1-\nu)}{3^{2\nu+1}\nu \Gamma(1+\nu) } (xx')^{\alpha+1/2} +  (xx')^{\alpha+1/2} \frac{\Lambda^\alpha}{2^{2\alpha+1} \alpha \Gamma^2(\alpha+1)} \ .
\end{align}
Plugging this into our expression for ${\cal{M}}_N$, we get
\begin{equation}
{\cal{M}}_N =  -\frac{2^{2\alpha + 1} \alpha\Gamma(\alpha+1)A_0^{\nu}}{\Gamma(1-\nu)}  \left[-\frac{\Gamma(1-\nu)}{3^{2\nu+1}\nu \Gamma(1+\nu) } +  \frac{\Lambda^\alpha}{2^{2\alpha+1} \alpha \Gamma^2(\alpha+1)}\right] \ .
\end{equation}
Adding this to the large-$k$ contribution, ${\cal{M}}_R$, we see that the $\Lambda$ dependence cancels out and gives our final result
\begin{equation}
{\cal{M}}_\nu = \frac{2^{2\alpha-1}\Gamma(\alpha)  }{3^{2\nu-1}\Gamma(\nu)}A_0^{\nu} \ .
\end{equation}
This result was also derived in Ref. \cite{PRL3}, by examining the statistics of a recurrent process of excursions, which give rise to a L\'evy walk. It can be seen to be consistent with our above results for the case of $\nu=1$, where it agrees with the first moment result, and for the case of $\nu=2$ for the appropriate second moment result.  It also agrees with the Airy Distribution result corresponding to the 1/3 moment~\cite{Crandall,Flajolet}.

\section{Conclusions}
In this paper, we have constructed what we call the Bessel Distribution, $P(A)$, namely the distribution of the area under a Bessel excursion.  The distribution is characterized by the parameter $U_0$, denoting the strength of the bias towards the origin. This corresponds to a  $d=1-U_0$ dimensional  Brownian random walk whose radius is described by a Bessel process. The case $U_0=0$ gives the well-studied Airy Distribution. The distribution is symmetric in the parameter $U_0+1$, and is nonanalytic at
$U_0=-1$, corresponding to $d=2$. The analytic continuation of the $U_0>-1$ distribution approaches a (cut-off) one-sided L\'evy $\alpha$-stable distribution with index $2/3$ and scale factor proportional to $(4-d)^{3/2}$ as  $d$ approaches 4 from below (i.e., $U_0$ approaches $-3$ from above).  

We have calculated the first and second moments of the Bessel Distribution as a function of $U_0$.  For the first moment, we obtained a closed-form answer, whereas for the second moment we obtain the result in terms of an infinite sum over generalized hypergeometric functions. We also found a closed-form result for the $\nu=(U_0+1)/3$ moment. In comparing to the $U_0=0$, Airy Distribution case, we obtain a novel form for the Airy Distribution directly in terms of the Airy function and its derivative.  We also demonstrate an alternative formalism for calculating the integer moments of the Airy Distribution.

It is clear that similar progress can be made on the distribution of the area under Bessel meanders, where the Bessel process is not constrained to return to the origin.  We are currently working in this direction, in particular as it has implication for the diffusion of cold atoms~\cite{PRL3}.

\ack
This work is supported in part by the Israel Science Foundation.
\bibliographystyle{apt}
\bibliography{apt}

\begin{thebibliography}{10}

\bibitem{AbS}
{\sc Abramowitz, M. and Stegun, I.~A.} (1964).
\newblock {\em Handbook of Mathematical Functions}.
\newblock National Bureau of Standards, Washington.

\bibitem{Bender}
{\sc Bender, C.~M. and Orszag, S.~A.} (1999).
\newblock {\em Advanced Mathematical Methods for Scientists and Engineers I}.
\newblock Springer-Verlag, New York.

\bibitem{Carmi}
{\sc Carmi, S. and Barkai, E.} (2011).
\newblock Fractional {F}eynman-{K}ac equation for weak ergodicity breaking.
\newblock {\em Physical Review E\/} {\bf 84,} 061104.

\bibitem{Crandall}
{\sc Crandall, R.~E.} (1996).
\newblock On the quantum zeta function.
\newblock {\em Journal of Physics A: Mathematical and General\/} {\bf 29,}
  6795--6816.

\bibitem{Darling}
{\sc Darling, D.~A.} (1983).
\newblock On the supremum of a certain {G}aussian process.
\newblock {\em The Annals of Probability\/} {\bf 11,} 803--806.

\bibitem{Flajolet}
{\sc Flajolet, R. and Louchard, G.} (2001).
\newblock Analytic variations on the airy distribution.
\newblock {\em Algorithmica\/} {\bf 31,} 361--377.

\bibitem{Hu}
{\sc Hu, Y. and Shi, Z.} (1997).
\newblock Extreme lengths in {B}rownian and {B}essel excursions.
\newblock {\em Bernoulli\/} {\bf 3,} 387--402.

\bibitem{Ito}
{\sc {It$\hat{\textrm{o}}$}, K. and H.~P.~McKean, J.} (1974).
\newblock {\em Diffusion Processes and their Sample Paths} vol.~125 of {\em
  Grundlehren der mathematischen Wissenschaften}.
\newblock Springer-Verlag, Berlin.

\bibitem{review}
{\sc Janson, S.} (2007).
\newblock Brownian excursion area, {W}right's constants in graph enumeration,
  and other {B}rownian areas.
\newblock {\em Probability Surveys\/} {\bf 4,} 80--145.

\bibitem{Kac}
{\sc Kac, M.} (1949).
\newblock On distributions of certain {W}iener functionals.
\newblock {\em Transactions of the American Mathematical Society\/} {\bf 65,}
  1--13.

\bibitem{PRL3}
{\sc Kessler, D.~A. and Barkai, E.} (2012).
\newblock Theory of fractional {L\'e}vy kinetics for cold atoms diffusing in
  optical lattices.
\newblock {\em Physical Review Letters\/} {\bf 108,} 203602.

\bibitem{Louchard}
{\sc Louchard, G.} (1984).
\newblock Kac's formula, {L\'e}vy's local time and {B}rownian excursion.
\newblock {\em Journal of Applied Probability\/} {\bf 21,} 479--499.

\bibitem{Majumdar}
{\sc Majumdar, S.~N. and Comtet, A.} (2005).
\newblock Airy distribution function: {F}rom the area under a {B}rownian
  excursion to the maximal height of fluctuating interfaces.
\newblock {\em Journal of Statistical Physics\/} {\bf 119,} 777--826.

\bibitem{PREBessel}
{\sc Martin, E., Behn, U. and Germano, G.} (2011).
\newblock First-passage and first-exit times of a {B}essel-like stochastic
  process.
\newblock {\em Physical Review E\/} {\bf 83,} 051115.

\bibitem{Penson}
{\sc Penson, K.~A. and {G\'orska}, K.} (2010).
\newblock Exact and explicit probability densities for one-sided {L\'e}vy
  stable distributions.
\newblock {\em Physical Review Letters\/} {\bf 105,} 210604.

\bibitem{Takacs}
{\sc {Tak\'acs}, L.} (1991).
\newblock A {B}ernoulli excursion and its various applications.
\newblock {\em Advances in Applied Probability\/} {\bf 23,} 557--585.

\bibitem{Zolotarev}
{\sc Zolotarev, V.~M.} (1961).
\newblock Expression of the density of a stable distribution with exponent
  $\alpha$ greater than one by means of a frequency with exponent $1/\alpha$.
\newblock {\em Selected Translations in Mathematical Statistics and
  Probability\/} {\bf 1,} 163--167 [Translation of {\em Dokl. Akad. Nauk SSSR}
  {\bf 98}, 735--738 (1954)].

\end{thebibliography}

 \end{document}